\documentclass[a4paper,11pt]{article}
\usepackage{pos}
\usepackage[style=numeric,sorting=none]{biblatex}
\usepackage{subcaption}
\usepackage{graphicx}
\smallskipamount

\title{Grand unification, small vs large representations, hadron colliders and all that }

\author*[a,b]{Goran Senjanovi\'c}
\author[c]{Michael Zantedeschi}

\affiliation[a]{Arnold Sommerfeld Center, Ludwig-Maximilians University,\\
  Munich, Germany}

\affiliation[b]{International Centre for Theoretical Physics,\\
Trieste, Italy}

\affiliation[c]{Tsung-Dao Lee Institute and School of Physics and Astronomy,\\
Shanghai Jiao Tong University,\\ Shanghai, China}

\emailAdd{goran.senjanovic@physik.uni-muenchen.de}
\emailAdd{zantedeschim@sjtu.edu.cn}

\abstract{We review minimal realistic grand unified models based on $SU(5)$ and $SO(10)$ gauge groups. The models with small Higgs representations and higher dimensional operators - under the assumption of no cancellations in proton decay amplitudes - predict nearby oases with new light particles. Two of them stand out: real fermion and scalar weak triplets. The former generates dynamically neutrino Majorana mass through the so-called type III seesaw mechanism, while the latter naturally gets a small vacuum expectation value and thus generically modifies $W$-boson mass. On the contrary, the renormalisable versions of these theories fail to determine the particle spectra. 
In particular, in the renormalisable version of the $SO(10)$ theory the mass of the right-handed gauge boson - normally assumed (and allowed) to lie far in the desert - could be even accessible at the LHC. Last but not least, we show that in the minimal supersymmetric $SU(5)$ theory all the superpartners can lie orders of magnitude above the weak scale, unless one requires strict naturalness. Thus, the so-called split supersymmetry does not imply any light states. 
}

\addtocontents{toc}{\setcounter{tocdepth}{4}} 
\FullConference{%
  Corfu Summer Institute 2022 "School and Workshops on Elementary Particle Physics and Gravity",\\
  28 August - 1 October, 2022\\
  Corfu, Greece}

\tableofcontents

\begin{document}
\maketitle
\section{Introduction}

Grand unification of the Standard Model forces is one of the most appealing avenues for Beyond the Standard Model physics. In its core lie two profoundly exciting predictions, the existence of magnetic monopoles and the proton decay, both desperately searched for decades. It has a potential capacity to predict its own scale from unification constraints, and in principle even the weak mixing angle. Its original appeal suffered from, or so it seemed in the early days, the prediction of the desert all the way to enormous energies, some 13-14 orders of magnitude above the weak scale~\cite{Georgi:1974yf}. So how in the world could it manifest itself at colliders? 

The answer is simple. While it was known that in general there could be an occasional oasis in the desert, it came as a great surprise that in some minimal theories, the oasis would have to lie at nearby energies and even be potentially accessible at the next hadron collider, if not at the LHC itself. In particular this happens in a minimal realistic extension~\cite{Bajc:2006ia} of the original $SU(5)$ theory~\cite{Georgi:1974sy} and even in the very minimal $SO(10)$ theory~\cite{Fritzsch:1974nn} based on small Higgs representations~\cite{Preda:2022izo}. In both cases, the original models must be augmented by higher-dimensional operators in order to account for realistic fermion spectra, but this can be done in a consistent and predictive manner. This is the topic of our short review.

The next section is devoted to the search for a realistic $SU(5)$ theory. We start by overviewing the situation in the  minimal model of Georgi and Glashow~\cite{Georgi:1974sy}, 
which unfortunately fails for a number of reasons: it does not unify the gauge couplings, it predicts wrong down quark-charged lepton mass relations and, just like the Standard Model, leads to massless neutrinos. It is a great pity since this theory has a unique capacity to predict all the nucleon decay branching ratios. The first two failures can be addressed by adding higher-dimensional operators, but the solution comes at the expense of  maximally rotating away proton decay, which requires severe fine-tuning. 

In any case, the masslessness of neutrinos requires going beyond the minimal theory. This is naturally achieved through the seesaw mechanisms~\cite{Minkowski:1977sc,Mohapatra:1979ia,Yanagida:1979as,GellMann:1980vs,Glashow:1979nm}, and we cover two minimal extensions based on type II and III seesaw. The latter case is of a particular interest since it predicts a light electromagnetically neutral weak fermion triplet, which generates neutrino mass and is potentially accessible at LHC. Moreover, it also suggests the existence of a light neutral scalar weak triplet, that generically leads to a deviation of the $W$-boson mass from its Standard Model value. 
Its decay rates into Standard Model degrees of freedom are uniquely determined by its mass and its vacuum expectation value that induces a deviation of $W$-mass from its Standard Model value~\cite{Senjanovic:2022zwy}. This, in principle, would fit nicely if the recent $W$-mass anomaly announced by the CDF collaboration~\cite{CDF:2022hxs} were true. However, even if CDF were wrong, this would still be of interest since it is a quantitative question that could show up before or later.

These predictions may not sound surprising; after all, it has been known for decades that low-energy supersymmetry accounts elegantly for the gauge coupling unification~\cite{Dimopoulos:1981dw,Einhorn:1981sx,Marciano:1981un}. However, it should be stressed that in the case of the type III seesaw, these are genuine predictions from the first principles, while low-energy supersymmetry is only a possibility, and is necessarily tied to the issue of naturalness since it provides a protection mechanism to keep the Higgs mass small in perturbation theory. Once one gives up the naturalness criterion, all hell breaks loose and we show that the scale of supersymmetry could be easily enormous, orders of magnitude above the weak scale.

In section \ref{so10} we discuss $SO(10)$ grand unified theory (GUT)~\cite{Fritzsch:1974nn} which naturally leads to small neutrino mass due to its structure and ties it to successful unification of gauge couplings through an intermediate mass scale~\cite{Preda:2022izo}. This intermediate scale was for a long time believed to be huge, not far from the unification scale, and was associated with a picture of a desert . However, in the minimal version based on small Higgs representations, it turns out that there must be an oasis at low energies. 
Section \ref{summary} is finally devoted to our conclusions and outlook.

\section{SU(5) theory}
\label{su5}

Although the minimal theory is ruled out, it is worthwhile to discuss its failures in order to pave the way to its minimal realistic extensions. 

\paragraph*{Minimal Model} Thus we start first with the original minimal $SU(5)$ theory introduced by Georgi and Glashow~\cite{Georgi:1974sy} which consists of three fermion generations placed in $\overline 5_{\rm F}$ and $10_{\rm F}$ representations, and the scalar adjoint $24_{\rm H}$ and fundamental $5_{\rm H}$ representations. We recollect here only the salient features of the model, for a more pedagogical {\it expose} the reader can consider e.g.~\cite{Senjanovic:2011zz}. The adjoint Higgs is responsible for the GUT scale symmetry breaking $M_{\rm GUT}$, whereas the fundamental one provides the electro-weak scale $M_{\rm W}$, with 
\begin{equation}
	\label{eq:vevs}
\begin{split}
\langle 24_{\rm H}\rangle &= M_{\rm GUT} \,{\rm{diag}}\left(1,1,1,-3/2,-3/2 \right); \\
 &\langle 5_{\rm H}\rangle^T = M_{\rm W} \left(0,0,0,0,1\right).
 \end{split}
\end{equation}
Notice that $\langle 5_{\rm H}\rangle$ accidentally preserves an $SU(4)$ symmetry between down quarks and charged leptons, only broken by $\langle 24_{\rm H}\rangle$. 
This will turn out to play a central role in fermion mass spectra.

\begin{figure}[t]
    \centering
    \includegraphics[width=0.6\textwidth]{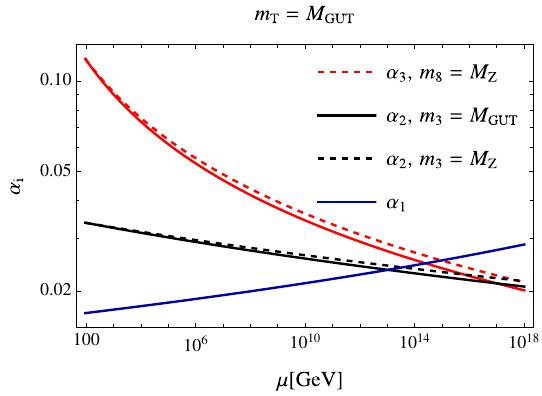}
    \caption{The failure of $m_3$ and $m_8$ at unifying the gauge couplings. }
    \label{fig:m3m8}
\end{figure}

The high-energy breaking leads to massive gauge bosons $X,Y$ which are responsible for proton decay. The associated lifetime is roughly given by 
\begin{equation}
\label{eq:plifetime}
	\tau_{\rm p}\simeq \frac{M_{\rm GUT}^4}{\alpha_{\rm GUT}^2}\, m_{\rm p}^{-5},
\end{equation}
where $m_{\rm p}$ is the proton mass and $\alpha_{\rm GUT}$ is the value of gauge couplings at $M_{\rm GUT}$. 
The most stringent bound on proton lifetime from Super-Kamiokande is about $\tau_{\rm p}\gtrsim10^{34}\,\rm{yrs}$ for the channel ${\rm p }\rightarrow \pi^0\, e^+$~\cite{Super-Kamiokande:2020wjk}. Typical values from the running are $\alpha_{\rm GUT}\simeq 1/40$  therefore obtaining
\begin{equation}
\label{eq:mgutnaive}
M_{\rm GUT}\gtrsim 4\cdot 10^{15}\,\rm{GeV}\,.
\end{equation}
The above bound is incompatible with the low-energy data as it can be seen from Fig.~\ref{fig:m3m8}.
The full-lines show the 1-loop running of the Standard Model couplings, clearly failing at unifying. Simply, the $U(1)$ gauge coupling meets the $SU(2)$ one too early.

Of course, this is not the end of the story since the model comes with new degrees of freedom on top of the Standard model ones, which, if light, could affect the running. Symmetry breaking, in fact, gives mass to the submultiplets contained in both $\langle 24_{\rm H}\rangle$ and $\langle 5_{\rm H}\rangle$. 
The former contains a scalar weak triplet $(1_C,3_W,0_Y)$ of mass $m_3$, ensuring the lightness of $W$ gauge boson and a color octet $(8_C,1_W,0_Y)$ of mass $m_8$, leaving the gluons massless, and a Standard Model singlet which is irrelevant from the point of view of gauge coupling running. The latter contains a scalar colored particle $(\overline{3}_C,1_W,1/3_Y)$ - a scalar version of the down quark - which must be heavier than about $10^{12}\rm GeV$ in order satisfy proton lifetime constrain \eqref{eq:plifetime}. The impact of this particle on the running can therefore be neglected. 
The effect of $m_3$ and $m_8$ is shown in Fig.~\ref{fig:m3m8} with dashed lines. As it can be seen, a light weak triplet helps unification but is still insufficient. 

Another failure of the model is that it predicts wrong low-energy Yukawa spectrum.
The renormalizable 
$d=4$ Yukawa interaction is given by
\begin{equation}
\label{eq:yukawas}
 \overline 5_{\rm F}\, Y_{\rm d} 10_{\rm F} \,5_{\rm H}^* + 10_{\rm F}  \,Y_{\rm u} 10_{\rm F} \,5_{\rm H} \, .
\end{equation}
Since the fermions get their masses from $\langle 5_{\rm H}\rangle$ - at least at the tree level - their spectra are not arbitrary. In particular, due to the unbroken $SU(4)$ symmetry, charged leptons and down quarks are predicted to have the same masses.
This, as it is well known, fails badly for the first two generations and even, albeit less pronounced for the third one.

So far the minimal model shows to be predictive, in fact, too predictive. It fails to reproduce both the realistic Standard Model fermion mass spectrum and the correct low-energy gauge couplings values.

\paragraph*{Higher-Dimensional Operators} A possible way out, without changing the particle content of the theory, is to allow for non-renormalizable operators. What happens is the following. Since $\langle 24_{\rm H}\rangle$ breaks the accidental $SU(4)$ of $\langle 5_{\rm H}\rangle^T$, the effective Yukawa couplings end up being arbitrary and so do fermion mass matrices, implying arbitrary unitary matrices that diagonalize them. The bad predictions are gone but this has a dramatic impact on proton decay. In particular, these unitary matrices can be highly non-diagonal resulting in the uncorrelation between proton lifetime and $M_{\rm GUT}$~\cite{Nandi:1982ew}. For example, $M_{\rm GUT}\sim 10^{14}\,\rm GeV$ could still be compatible with proton lifetime under suitable tuning of these matrices~\cite{Dorsner:2004xa}. 

The presence of higher-dimensional operators can also address the lack of gauge coupling unification by changing the boundary conditions at $M_{\rm GUT}$ via operators such as~\cite{Shafi:1979qb}
\begin{equation}
	\label{eq:higher}
	\mathcal{O}= \frac{1}{\Lambda^n}\,{\rm Tr}\,F_{\mu \nu} \langle 24_{\rm H}\rangle^n F^{\mu \nu},
\end{equation}
where $n\geq 1$. A cutoff $\Lambda \gtrsim 10\,M_{\rm GUT}$ suffices to guarantee gauge coupling unification, while keeping the $SU(5)$ symmetry well defined. For the sake of simplicity, we include in our discussion only the $5$-dimensional operator corresponding to $n=1$. In the vacuum, from Eq.~\eqref{eq:higher} the unification conditions is modified to~\cite{Shafi:1979qb}
\begin{equation}
\label{eq:gaugeunif}
	\left(1-\epsilon\right)\alpha_3 (M_{\rm GUT}) =\left( 1+\frac{3}{2} \epsilon \right)\alpha_2(M_{\rm GUT})
							       = \left( 1+\frac{1}{2}\epsilon \right) \alpha_1(M_{\rm GUT}).
\end{equation}
where $\epsilon =M_{\rm GUT}/\Lambda$. The sign of $d=5$ term in \eqref{eq:higher} is arbitrary, and for the unification to work, $\epsilon$ must be positive, so that \eqref{eq:gaugeunif} can accommodate the problem of $\alpha_1$ being too big at the GUT scale. While this is achieved by choice, it automatically ensures that the $\alpha_3$ coupling remains large enough and it does so in the physically allowed region $10\, M_{\rm GUT} \lesssim \Lambda \ll M_{\rm pl}$. 
This can be seen from the unification condition
\begin{equation}
\label{eq:mgutepsilon}
	\frac{M_{\rm GUT}}{M_{\rm Z}} =  \exp\left\{ \frac{\pi}{21}\left[5\left(1+\frac{1}{4}\epsilon \right)\alpha_1^{-1} - 3 	\left( 1 + \frac{5}{4}\epsilon \right) \alpha_2^{-1} - 2\left( 1- \frac{5}{4}\epsilon \right)\alpha_3^{-1}\right]\right\}
	 \left(\frac{M_{\rm Z}^2}{m_3\, m_8} \right)^{\frac{1}{42}},
\end{equation}
where the gauge couplings are evaluated at $M_{\rm Z}$ and $\epsilon$ turns out to be in the range $10 \lesssim \epsilon \lesssim 100$. For light enough $m_3,m_8$, unification can be achieved, with $M_{\rm GUT}\gtrsim 10^{14}\,\rm GeV$.

While well defined, this approach admittedly abandons minimality - one must appeal to new physics above the cutoff $\Lambda$, whether stemming from quantum gravitational effects or new heavy fields. The alternative of large scalar representations that can work at the renormalizable level brings, however, huge threshold effects and prevents the study of the central aspect of GUT - the study of the unification of gauge couplings. In view of this, it is worthwhile to explore the small representation approach, which is the main focus of this work.

Indeed, in the quest for a minimal viable realization of the GUT framework based on small representations, it would be desirable to have a theory for neutrino mass - a right-handed neutrino is missing in the above example. Moreover, a second item in the whishlist is the predictions of new phenomena in the low-energy sector of the theory. 

Within $SU(5)$ there are two particularly attractive alternatives, with minimal changes of the theory and predictive outcomes, both obtained by the addition of a single field augmented with higher-dimensional operators as in the original minimal theory. This is the price that one has to keep paying for sticking to minimality. Of course, one can make the theory renormalizable by adding more fields, but here we shy away from that. The two models discussed below are motivated by the dynamics of the seesaw mechanism~\cite{Minkowski:1977sc,Mohapatra:1979ia,Yanagida:1979as}, with possibly observable consequences.

Since we have included higher-dimensional operators to cure the charged-fermion mass relations, it would seem natural to appeal to $d=5$ effective operator of the Weinberg type~\cite{Weinberg:1979sa}:
\begin{equation}
    \mathcal{O}= \frac{1}{\Lambda}\left(\overline{5}_{\rm F}\, 5_{\rm H} \right)\left(\overline{5}_{\rm F}\, 5_{\rm H} \right),
\end{equation}
implying neutrino mass of order $m_\nu \simeq M_{\rm W}^2/\Lambda$.
Since the cutoff must be large, $\Lambda \gtrsim 10\, M_{\rm GUT} \gtrsim 10^{16}\,$GeV, neutrino mass would end up far too small $m_\nu \lesssim 10^{-3}\,$eV - clearly, more is needed.

One could in principle just add the $SU(5)$-singlet fermions and appeal to the so-called type I seesaw, but this would be of no help for the unification of gauge couplings, if one does not fine-tune proton decay. Moreover, it would completely lack any predictivity and in a sense would not even be a minimal extension, since at least two such states are needed. 
Instead, it is more natural and more predictive to opt for the addition of a single multiplet to the original theory. As we show now, this amounts to other versions of the seesaw mechanism, the so called type II and type III.

\paragraph*{Neutrino mass: type II seesaw}
 Instead of fermion singlets, the right-handed neutrinos, one may opt for the type II 
 seesaw~\cite{Magg:1980ut,Mohapatra:1980yp,Lazarides:1980nt} through the addition of a symmetric scalar $15_{\rm H}$ field (for a recent study, see~\cite{Dorsner:2005fq}). This introduces a new Yukawa coupling
\begin{equation}
\label{eq:yukawaII}
 \overline 5_{\rm F}\, Y_\nu 15_{\rm H} \, \overline 5_{\rm F}\, ,
\end{equation}
which then induces neutrino Majorana mass through a small vacuum expectation value of $15_{\rm H}$, obtained from a tadpole term in the potential%
\begin{equation}
\label{eq:tadpole15}
 \mu\, 5_{\rm H} 15_{\rm H}^*\, 5_{\rm H}\, .
\end{equation}
This gives $m_\nu \simeq \langle 15_{\rm H} \rangle \simeq \mu \, M_{\rm W}^2/m_{15}^2$, perfectly consistent with data since both $\mu$ and the mass $m_{15}$ of $15_{\rm H}$ are arbitrary parameters.
While it may be considered somewhat an ad-hoc outcome of model building, it has an appeal of being potentially verifiable as the origin of neutrino mass. Namely, the weak triplet in $15_{\rm H}$ contains doubly charged scalars, whose branching ratio into charged leptons are dictated by the neutrino mass matrix~\cite{Garayoa:2007fw}. 

The problem with this simple model is the unification of gauge couplings. It implies a too low unification scale~\cite{Dorsner:2005fq}, which would require judicious cancellations in proton decay amplitudes. Since we are shying away from this option, we rather turn to an alternative, equally simple model.

\paragraph*{Neutrino mass: type III seesaw}  
This extension of the minimal $SU(5)$,  instead of $15_{\rm H}$, is based on a fermionic adjoint $24_{\rm F}$ field~\cite{Bajc:2006ia}, that contains, among other components, a weak triplet and singlet. The neutral component of the triplet plays a role analogous to a singlet (the right-handed neutrino) and leads to the type III seesaw~\cite{Foot:1988aq}. Thus, we have a combination of both type I and type III with a prediction of one massless neutrino. In this, the crucial role is played by the neutrino Dirac Yukawa coupling
\begin{equation}
\label{eq:yukawaIII}
 \overline 5_{\rm F}\, Y_{\rm D}\, 24_{\rm F} \, 5_{\rm H}\, .
\end{equation}
Together with the mass term $24_{\rm F}\, 24_{\rm F}$, analog of right-handed neutrino mass, one reproduces the usual seesaw formula for neutrino mass. The essential point of the type III aspect is that it can be direclty verified at hadron colliers if the weak triplet fermion is light enough.

Given the new degrees of freedom, unification can be attained without resorting to operators such as \eqref{eq:higher} as it can be seen from the unification conditions
\begin{equation}
\label{eq:mgut24f}
	\frac{M_{\rm GUT}}{M_{\rm Z}} = \exp \left\{ \frac{\pi}{21}\left[5\alpha_1^{-1} - 3  \alpha_2^{-1} - 2\alpha_3^{-1}\right]\right\}
	 \left(\frac{m_{\rm LQ}^4\,M_{\rm Z}}{m_3^{1/2}\, m_8^{1/2} m_{3\rm F}^2\, m_{8\rm F}^2} \right)^{\frac{1}{21}},
\end{equation} 
where the labels $3 \rm F$, $8\rm F$ and $\rm LQ$ denote respectively the weak triplet, the colour octet and the leptoquark component of $24_{\rm F}$ (the corresponding scalar leptoquark sub-multiplet is massless as it is eaten by the $X,Y$ gauge bosons). 
In the previous section, a low $m_3$ and $m_8$ could increase unification up to $10^{14}\,\rm{GeV}$ (c.f., \eqref{eq:mgutepsilon}). It is not surprising then that lowering also their fermionic counterpart can give $M_{\rm GUT}\gtrsim 4\cdot  10^{15}\,\rm GeV$, compatible with proton lifetime.

Assuming no cancellations in proton decay amplitudes, the above requirement imposes the presence of the weak triplet fermions and scalars close to the reach of the LHC as it can be easily checked from a direct inspection of \eqref{eq:mgut24f}.
While the phenomenology of the scalar will be discussed below in Sec.~\ref{Wboson}, its fermionic counterpart offers the exciting prospect of probing the see-saw mechanism at present-day energies. For details on its signatures see~\cite{Bajc:2007zf,Arhrib:2009mz}. Meanwhile, LHC has raised the lower limit on its mass to roughly $\rm TeV$~\cite{Novak:2020jju},   unification constraints impose an upper bound on the order of  $10\,\rm TeV$.

Notice that from the point of view of gauge couplings unification, the addition of the fermionic degrees of freedom is equivalent to the gauginos of the Supersymmetric Standard Model, which is known to unify provided the supersymmetry breaking scale is around $\rm TeV$ energies. However supersymmetry, when embedded in a grand unified framework, due to a proliferation of states 
makes no prediction about the energy scale of its spectrum as we now show. 

\subsection*{W-boson mass}
\label{Wboson}
The gauge coupling unification, as we said, hints at a rather light scalar weak triplet $3_{\rm H}$. This turns out to emerge in both also in the $SO(10)$ model based on small representations~\cite{Preda:2022izo}, as shown in Sec.~\ref{so10}. The lower limit on its mass follows from LHC searches $m_{3}\gtrsim 250\,\rm GeV$~\cite{Chiang:2020rcv}.
It has been known~\cite{Buras:1977yy} for a long time that due to the tadpole interaction with the Standard Model Higgs doublet $\Phi$
\begin{equation}
\label{eq:mu}
\mu\, \Phi^{\dagger}3_{\rm H}\,\Phi,
\end{equation}
the triplet acquires a non-vanishing vacuum expectation value 
\begin{equation}
\label{eq:v3}
	\langle 3_{\rm H}\rangle \simeq \frac{\mu}{4\pi \alpha_2}\left(\frac{M_{\rm W}^{\rm SM}}{m_3} \right)^2,
\end{equation}
where $M_{\rm W}^{\rm SM}$ is the Standard Model contribution to $W$-mass. Traditionally, such contribution has been neglected due to the common belief that this particle should lie at the GUT scale. On the contrary, as we have shown, it is precisely this particle that generically needs to be light to ensure unification. 
In turn, this leads to a mixing angle $\theta$ between $3_{\rm H}$ and $\Phi$, which within grand unification, is determined mainly by operator \eqref{eq:mu}~\cite{Senjanovic:2022zwy}, therefore leading to
\begin{equation}
\label{eq:theta}
	\theta \simeq \sqrt{4\pi \alpha_2} \,\frac{\langle 3_{\rm H} \rangle}{M_{\rm W}^{\rm  SM}}, 
\end{equation}
which notably leads to a deviation of $W$-mass from its Standard Model predicted value
\begin{equation}
\label{eq:mwmass}
	M_{\rm W}^2\simeq \left(M_{\rm W}^{\rm SM}\right)^2\left(1 + \theta^2\right)
\end{equation}
while leaving $Z$-boson mass unchanged. 
This could explain the CDF-collaboration measurement of $W-$boson mass $M_{\rm W}=80.433 \pm 0.009 \, \rm GeV$~\cite{CDF:2022hxs}, in conflict with the Standard Model value $M_{\rm W}^{\rm SM}=80.357 \pm 0.006\,\rm GeV$. For $\sqrt{4\pi \, \alpha_2(M_{\rm W})}= 0.6517$~\cite{ParticleDataGroup:2020ssz}, Eq.~\eqref{eq:mwmass} implies $\langle 3_{\rm H}\rangle\simeq 4\, \rm GeV$, or $\theta\simeq 0.04$.
It can be easily checked that, for $\langle 3_{\rm H}\rangle \sim \mathcal{O}(\rm GeV)$, electroweak spontaneous symmetry breaking stability requires the triplet to lie below $10 \,\rm TeV$~\cite{Senjanovic:2022zwy}.

Notice though that the $\mu$ coupling is self-protected, so that it can be naturally arbitrarily small. In other words, one cannot claim a detectable $W$-mass deviation.

Naively, any theory with the ad-hoc addition of a scalar triplet, even one based on large representations, can accommodate a light triplet. However, in the context of grand unification, no phenomenological input regarding $W-$mass deviation from the Standard Model value is needed. Moreover, in the case of minimal $SU(5)$ and its extension with $24_{\rm F}$ discussed above~\cite{Bajc:2006ia}, it is the only possible source for such deviation. In models with a type II seesaw, radiative corrections could in principle address the CDF-deviation~\cite{Cheng:2022jyi,Heeck:2022fvl}. This makes for a rather entangled situation for the minimal extensions based on the addition of a $15_{\rm H}$~\cite{Dorsner:2005fq}, where multiple sources could change $W$-boson mass.  

In this sense, the $SU(5)$ model of~\cite{Bajc:2006ia}, with an adjoint fermion, stands out in terms of its predictivity. The only light scalar besides the SM Higgs is the weak triplet, and, as a consequence, its decay rates into Standard Model degrees of freedom are determined uniquely by $W-$mass deviation at leading order. The decays into the bosonic final states are given by~\cite{Senjanovic:2022zwy}, 
\begin{equation}
\begin{split}
\label{eq:charged}
  &\Gamma(H^0\rightarrow W^+W^-)\simeq 2\Gamma(H^0\rightarrow ZZ)\simeq2\Gamma(H^0\rightarrow h^0h^0) \\
   &\simeq \Gamma(H^+\rightarrow W^+Z)\simeq \Gamma(H^+\rightarrow W^+h^0)
   \simeq \frac{\alpha_2\theta^2}{16} \frac{m_3^3 }{M_{\rm W}^2}
    \qquad
    \end{split}
\end{equation}
where $H^0$ and $H^+$ denote respectively the neutral and charged component of $3_{\rm H}$ and $h^0$ is the neutral component of the Higgs doublet $\Phi$.
The decays into fermions are highly suppressed due to small fermion masses - except for the top quark final state for the sufficiently heavy triplet. In this case, one gets~\cite{Senjanovic:2022zwy}
\begin{equation}
\label{eq:neutral}
 \Gamma(H^0\rightarrow  t \overline{t})\simeq  \Gamma(H^+\rightarrow t\overline b)\simeq \frac{3\alpha_2\theta^2}{8} \frac{m_{\rm t}^2\, m_3}{M_{\rm W}^2},
\end{equation}
where $m_{\rm t}$ denotes top mass. Eqs.~(\ref{eq:charged}) and (\ref{eq:neutral}) are exact only in the limit of massless final states, and provide a good approximation for $m_3\gtrsim 500\,\rm GeV$. For lighter $m_3$ the exact decay rate can be found in our previous work~\cite{Senjanovic:2022zwy}.

\subsection*{Low-energy supersymmetry and grand unification}
Let us briefly recall the rise (and fall?) of low-energy supersymmetry. Ordinary grand unified theories are often said to be plagued by two serious issues: the perturbative lightness of the Higgs boson and the necessary fine-tuning of its mass against $M_{\rm GUT}$. As well known, the Higgs mass term is quadratically sensitive to large UV scales, such as the unification one or the Planck scale. This makes its relative smallness unnatural in the technical sense~\cite{tHooft:1979rat} - namely, the quantum loop effects are much larger than the experimental value. This {\it per se} would not be a problem - or even an issue - if it did not require fine-tuning between the tree-level and the radiative corrections. 

The low-energy supersymmetry cures this elegantly through the cancellations of quadratic divergences and thus protects the Higgs mass (or the weak scale) to all orders in perturbation theory. However, it must be kept in mind that the fine-tuning persists as much as before, at least in the context of the minimal grand unified models discussed here. This is the infamous doublet-triplet splitting problem, which requires the color triplet partner in $5_{\rm H}$ to lie close to the unification scale to be compatible with the observed proton lifetime.

The supersymmetric Standard Model partners then correct the above mentioned running, and do so at a scale $\Lambda^{\rm MSSM}$ not so far from current energies as it can be seen from the 1-loop renormalization-group condition 
\begin{equation}
\label{eq:lambdamssm}
	\Lambda^{\rm MSSM} = M_{\rm GUT}\left( \frac{M_{\rm GUT}}{M_{\rm Z}}\right)^{-\frac{11}{2}} \exp\left\{-\frac{\pi}{4} \left(2 \alpha_3^{-1} + 3 \alpha_2^{-1} - 5 \alpha_{1}^{-1}\right)_{M_{\rm Z}}\right\}.
\end{equation}
The combination of couplings in \eqref{eq:lambdamssm} is independent of the scale of the second Higgs doublet necessary for anomaly cancellation. 
\begin{figure}[t]
  \centering
  \begin{subfigure}[b]{0.48\linewidth}
    \includegraphics[width=\linewidth]{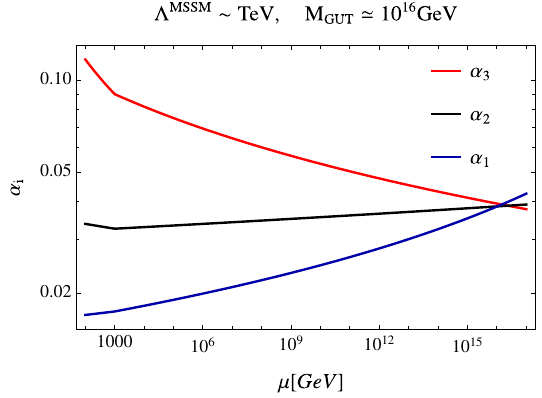}
  \end{subfigure}
  \begin{subfigure}[b]{0.5\linewidth}
    \includegraphics[width=\linewidth]{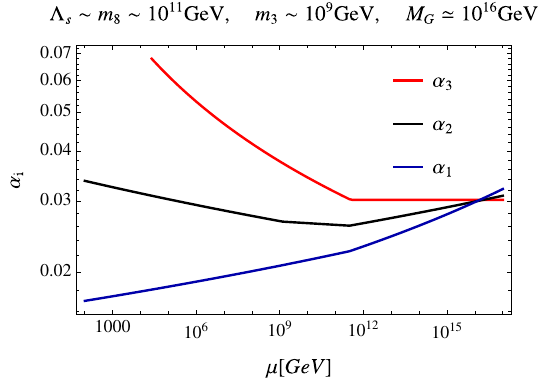}
  \end{subfigure}
  \caption{\textit{Left:} Standard Model running with super-partners around $\rm TeV$ scale. \textit{Right: } Minimal supersymmetric $SU(5)$ running. An example with high scale.}
  \label{fig:susy}
\end{figure}
Using $M_{\rm Z}= 91.19 \,\rm{GeV}$, $\alpha_3(M_{\rm Z})= 8.44^{-1}$, $\alpha_2(M_{\rm Z})= 29.57^{-1}$ and $\alpha_1(M_{\rm Z})= 59.02^{-1}$~\cite{ParticleDataGroup:2020ssz}, it can be seen from \eqref{eq:lambdamssm}, for $M_{\rm GUT}\sim 10^{15\div 16}\,\rm GeV$, $\Lambda^{\rm MSSM}$ ranges between $1\div 10 \,\rm TeV$ (see left panel in Fig.~\ref{fig:susy}).
We should stress that this was predicted~\cite{Dimopoulos:1981dw,Einhorn:1981sx,Marciano:1981un} years before LEP managed to measure correctly the weak mixing angle. Moreover, in~\cite{Marciano:1981un} it was argued that the top quark had to be extremely heavy, $m_t \simeq 200\,$GeV in order for the weak mixing angle to agree with the prediction from the low-energy supersymmetry, at the time when experiment seemed to indicate the opposite. Some quarter of century later, this turned out to be prophetically correct when the top was discovered. These two facts, the unification and heavy top quark
can be considered a success of low-energy supersymmetry.

Thus, the desire for naturalness seems to fit beautifully with the gauge coupling unification. One may wonder what happens if naturalness is given up - after all, we are still plagued by the fine-tuning as remarked above. 
Doing this analysis requires embedding the Supersymmetric Standard Model in grand unification. For simplicity and maximal predictivity, we focus here on the minimal supersymmetric $SU(5)$ 
model~\cite{Dimopoulos:1981zb}, implying that the adjoint $24_{\rm H}$ field now comes together with its fermionic superpatner.  To ease the reader's pain, in the following we assume both scalar and fermionic degrees of freedom to have the same mass, so $m_3$ and $m_8$ now denote superfields and thus their threshold impact is significantly larger.

It is easy to see that sfermions do not alter the unification conditions - as long as they stay roughly degenerate - so their effects can be simply ignored.
Now, Eq.~\eqref{eq:lambdamssm} becomes \cite{SZ}\footnote{We thank Borut Bajc for pointing out that this formula was already present in~\cite{Bajc:2015ita}, in the context of low energy supersymmetry. Here we wish to emphasise the fact the scale of supersymmetry may be at astronomical energies, when it comes to unification.}
\begin{equation}
\label{eq:lambdasm}
\Lambda = \Lambda^{\rm MSSM} \left( \frac{M_{\rm GUT}^2}{m_3 \,m_8}\right)^{3/4}.
\end{equation}
As it can be seen a weak-triplet or color octet just slightly lighter than the GUT scale can significantly change the supersymmetry scale. For example, in the right panel of Fig.~\ref{fig:susy} all new particle states are above $10^{9}\,\rm GeV$~\cite{SZ}. 

It is therefore clear that supersymmetry, due to the proliferation of states, makes no prediction about the scale of new physics. This is in sharp disagreement with~\cite{Arkani-Hamed:2004ymt}, that advocated giving up completely on naturalness but maintaining the unification of gauge couplings and arguing that gauginos must be light for the sake unification. This is wrong, as we have just shown. Once the naturalness is abandoned, the particles masses are free and thus one must allow for the threshold effects of $m_3$ and $m_8$. In short, any low-energy supersymmetric partner would be just wishful thinking without naturalness. On a positive side, as long as one does not ask for absolute naturalness and allows for small cancellations between the tree-level and loop contributions to the Higgs boson mass, low-energy supersymmetry is as healthy as it ever was, and it could be just around the corner. Rumors about its death are greatly exaggerated, to paraphrase Mark Twain.

Since supersymmetry is not forced on us by any consistency requirements and since it does not need to be accessible at colliders for a long time, 
it is more natural to go back to the non-supersymmetric case.

\section{SO(10) theory}
\label{so10}
As shown in the previous Sections, sticking to small representations provides, within Grand Unification, hints of new particle states near present day energies. This is already clear from the 1-loop renormalization group equation of Georgi-Glashow model in \eqref{eq:mgutepsilon}, where light particle states are desirable in order to be compatible with the observed proton lifetime. 

At the renormalizable level this theory is not phenomenologically viable, unless higher-dimensional operators taking care of the low-energy fermionic spectrum and of gauge coupling unification are included. However, the model lacks a right-handed neutrino, and therefore can not dynamically generate neutrino mass, tying new physics with it. 
Of course, such fermionic degrees of freedom could be added by hand as singlets. Still, this might be considered an unsatisfactory solution, and further model building might be seeked. 

We briefly described two existing extensions~\cite{Bajc:2006ia,Dorsner:2005fq} of Georgi-Glashow model based on small representations and higher-dimensional operators which realize the seesaw paradigm in order to ensure a viable fermionic spectrum. 
Indeed, introducing higher-dimensional operators implies the incompleteness of the model. The cutoff scale is, however, at least an order of magnitude away from $M_{\rm GUT}$, ensuring the applicability of the effective field theory. 

The advantage of focusing on these extensions based on small representations is that the longevity of proton lifetime (and therefore the requirement of a heavy $M_{\rm GUT}$) still hints at the presence of light particle states, precisely as in the minimal Georgi-Glashow realization, c.f., \eqref{eq:mgut24f}.
Of course, one could stick to renormalizable models by adding bigger multiplets. However, in such scenario particle thresholds would no longer be under control, making the spectrum completely arbitrary. An example of this was showed in the minimal supersymmetric grand unified model of the previous Section. 

As already stated several times, the main reason for extending the minimal $SU(5)$ model is to obtain a dynamical picture of neutrino mass. However, this becomes, in a sense, an {\it a posteriori, ad-hoc} construction, just in order to explain already established facts of nature. 
From this point of view, $SO(10)$ grand unification~\cite{Fritzsch:1974nn} can be viewed as a more natural candidate. Moreover, a family of fermions is united within a single $16_{\rm F}$ dimensional multiplet, which is a motivation {\it per se} to go beyond $SU(5)$. On top of this, $16_{\rm F}$ contains also a right-handed neutrino $\nu^c$
\begin{equation}
    16_{\rm F}= \left(q,u^c,d^c,l,e^c,{\color{red}\nu^c}\right)_{\rm L},
\end{equation}
where in obvious notation $q,l$ denote the quark and lepton weak doublets and $u^c,d^c,e^c$ the anti-quarks and positron weak singlets. 
As a consequence, $SO(10)$ provides a natural framework for the realization of the see-saw picture that leads naturally to small neutrino mass. For a pedagogical review of the subject, the reader is directed to~\cite{Senjanovic:2011zz}.  

An obvious question is whether in this theory some statements about its scales can be made, similarly to what happened for $SU(5)$. At a fist glance this is not the case due to the existence of an intermediate scale in which new gauge bosons significantly correct the running, therefore ensuring unification. 

\begin{figure}[t]
    \centering
    \includegraphics[width=0.6\textwidth]{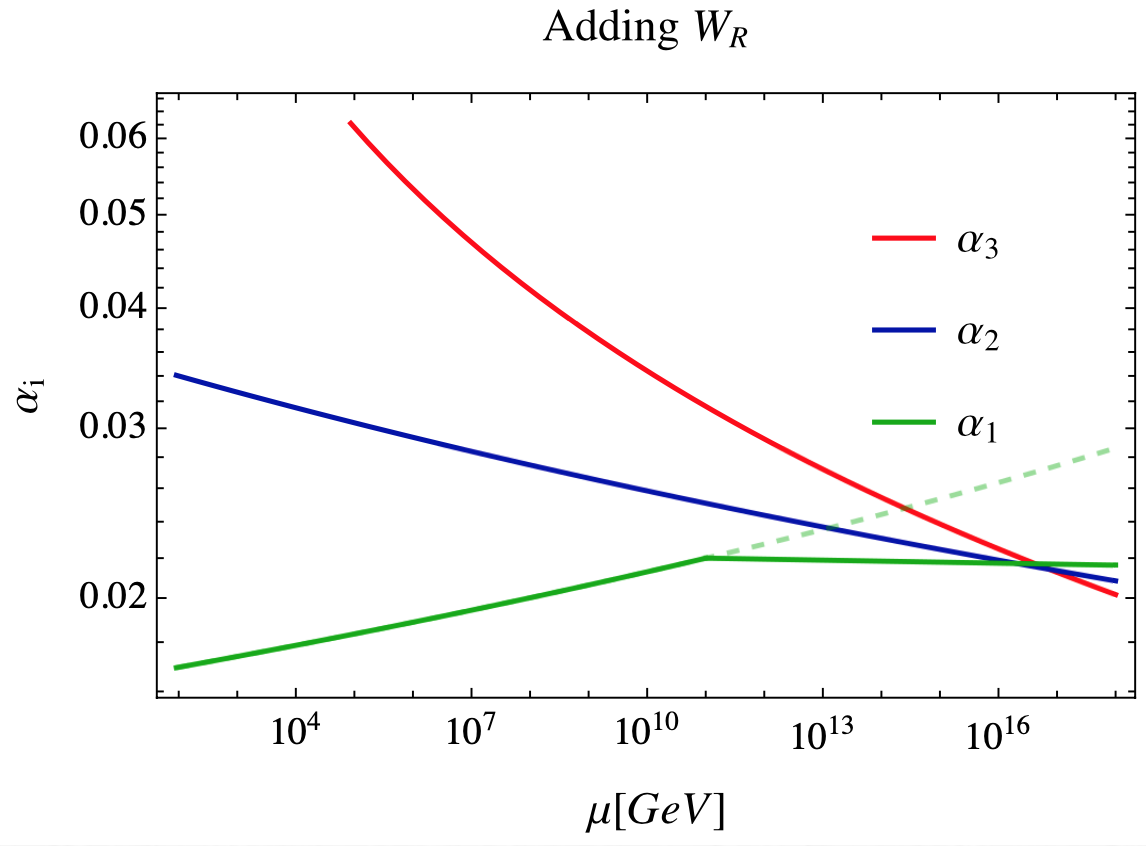}
    \caption{Effect of right-handed gauge coupling on the running.}
    \label{fig:LRrunning}
\end{figure}
To better appreciate this notice that $SO(10)$ contains the Left-Right symmetric group $SO(10)\supset SU(2)_{\rm L} \times SU(2)_{\rm R} \times U(1)_{\rm B-L} \times SU(3)_{\rm C}$, which then needs to be further broken to the Standard Model. This intermediate scale $M_{\rm I}$ is arbitrary and can be adjusted to ensure unification as it can be seen from Fig.~\ref{fig:LRrunning}. There, integrating in the right gauge boson $W_{\rm R}$ at around $M_{\rm I}\sim 10^{11}\,\rm GeV$ suffices to sufficiently slow down the $U(1)_{\rm Y}$ coupling (green-line) ensuring gauge coupling unification (the dashed-green line shows the Standard Model running).

The following picture emerges: Regardless of the symmetry breaking pattern - no matter whether it is Left-Right or Pati-Salam quark-lepton symmetry~\cite{Pati:1974yy} - the presence of an intermediate scale can always take care of gauge coupling unification and, as a consequence, no constraints on the particle spectrum of the model content arise. In particular, all new particle states can be much heavier than present-day energies which has been the conventional wisdom since the Early 80's~\cite{delAguila:1980qag,Rizzo:1981su} leading to a long-held belief of a desert from $M_{\rm W}$ up to $M_{\rm GUT}$, or $M_{\rm I}$ at best.

Surprisingly, this recently turned out not to be the case~\cite{Preda:2022izo}. We searched, with the same spirit used in $SU(5)$, for a minimal phenomenologically viable $SO(10)$ scenario based on small representations and higher-dimensional operators. We found a model requiring new particle states below $10\,\rm TeV$ the details of which we show below. 

The key-point is that, by sticking to small representations, the need for a neutrino mass value compatible with experiments forces the intermediate scale $M_{\rm I}$ close to $M_{\rm GUT}$. Practically, there is no intermediate scale in this model. This simple fact reintroduces, from the point of view of gauge coupling unification, the necessity of light particle states we observed in minimal $SU(5)$ and its extensions.

\subsection*{SO(10): a case for hadron colliders}
The minimal version of the theory (with small representations) besides three $16_{\rm F}$ spinors (fermion generations), contains the following Higgs scalars
\begin{equation}\label{higgs}
45_{\rm H}; \,\,\,\,\,  16_{\rm H}; \,\,\,\,\,  10_{\rm H}\,,
\end{equation}
where we use the notation that specifies the representation content. The $45_{\rm H}$ field is an adjoint, antisymmetric representation, which together with the spinorial $16_{\rm H}$ Higgs field, serves to break  the GUT symmetry to the Standard Model gauge symmetry. 
 Finally, a complex $10_{\rm H}$ vector is whether the Standard Model Higgs doublet resides, which then completes the breaking down to electromagnetic charge and color gauge invariance 

The truly minimal Higgs sector would use a real $10_{\rm H}$, but that would imply a single Yukawa coupling and a single vacuum expectation value, predicting all fermion masses being equal. In particular, top-bottom mass equality would  be impossible to cure since the higher-dimensional operators are simply not large enough. This is why  $10_{\rm H}$ must be complex, or in other words, one has two real $10_{\rm H}$'s, with two Yukawa couplings and two vacuum expectation values, which allows to split up from down quarks. 

While necessary, this is not sufficient. The Standard Model doublets in $10_{\rm H}$ are singlets under Pati-Salam quark-lepton symmetry, which implies the same masses for down quarks and charged leptons, clearly wrong. One could add more Yukawa Higgs scalars, but that would imply either $120_{\rm H}$ or/and $126_{\rm H}$ representation, against the original premise of employing small representations. Thus, the necessity of augmenting the Lagrangian with higher-dimensional operators. One has then the Yukawa sector, schematically written
\begin{equation}\label{yukawa}
{\cal L}_Y = 16_{\rm F}  \left (10_{\rm H} + 10_{\rm H} \frac {45_{\rm H}}{\Lambda} + \frac {16_{\rm H}^* 16_{\rm H}^*} {\Lambda}+  10_{\rm H} \frac {45_{\rm H}^2}{\Lambda^2} \right )16_{\rm F},
\end{equation}
where $\Lambda$ denotes the cutoff of the theory. In order for this expansion to be perturbatively valid, we take hereafter $\Lambda \gtrsim 10 \,M_{\rm GUT}$.
The first term provides the mass for the top quark. The second and fourth term induce an effective $120$ and $126$ effective coupling, sufficing to generate a realistic Yukawa spectrum. 

{\textit{Neutrino mass} }
Without the third term in \eqref{yukawa}, neutrino as other fermion would get the Dirac mass from the first two terms and its lightness would remain a mystery. This term, however, provides the mass for the right-handed neutrino N (the $SU(5)$ singlet heavy lepton), on the order of

\begin{equation}\label{Nmass}
m_{\rm N} \simeq \frac {\langle 16_{\rm H} \rangle^2}{\Lambda}\,.
\end{equation}
We should also mention the two-loop radiatively induced N mass~\cite{Witten:1979nr}, 
which is roughly given by $m_{\rm N} \lesssim (\alpha / \pi)^2 \langle 16_{\rm H} \rangle^2 / M_{\rm GUT}$. Now, there is an upper limit on the cutoff, naively the Planck scale, but in reality scaled down by the number of real physical states $n$ of the theory $\Lambda \lesssim M_{\rm Pl} / \sqrt n$~\cite{Dvali:2007hz,Dvali:2007wp}. Since $n \gtrsim 100$ in this theory, and $M_{\rm GUT} \gtrsim 10^{15}\, {\rm GeV}$ for the sake of proton's stability, one has $\Lambda \lesssim 100\, M_{\rm GUT}$. This implies that the dominant contribution to N mass comes from \eqref{Nmass}.

Once the Standard Model symmetry breaking is turned on, implying non-vanishing Dirac mass term $M_{\rm D}$, light neutrinos get their masses through the seesaw mechanism
\begin{equation}\label{seesaw}
M_\nu \simeq - M_{\rm D}^T \frac{1}{M_{\rm N}} M_{\rm D}\,.
\end{equation}
Due to the presence of higher-dimensional operators we lose control over $M_{\rm D}$ and $M_{\rm N}$, so it would appear that no predictions could be made. The point , however, is that the top quark gets the mass from the leading $d=4$ term in \eqref{yukawa}, implying then for the third generation
\begin{equation}\label{thirddirac}
m_{\rm D_3} = m_{\rm t}\,,
\end{equation}
where $m_{\rm D_3}$ is the third generation neutrino Dirac mass term. 
This innocent looking relation turns out to be the crux to it all. The argument is simple and is based on the failure of the minimal $SU(5)$ theory, as already mentioned in the previous Section. It seems to tell us that we ought to avoid the $SU(5)$ route or equivalently, the single step breaking scenario - but as we will see, we are being pushed into it. 

In other words, it is $\langle 45_{\rm H} \rangle$ that breaks the GUT symmetry down to the intermediate one, to be broken then by the $\langle 16_{\rm H} \rangle = M_{\rm I}$.  In turn, from \eqref{Nmass} and \eqref{thirddirac}, one has a potentially too large neutrino mass
\begin{eqnarray}\label{nueffect}
m_{\rm \nu} \simeq  \frac {(m_{\rm D_3})^2} {m_{\rm N}}  \simeq \frac {m_{\rm t}^2 \Lambda} {M_{\rm I}^2}\, .
\end{eqnarray}
The smallness of neutrino mass sets the stage then for the main predictions of the theory. The way unification was imagined to work all these years - in the usual picture of the desert in energies from $M_{\rm W}$ to $M_{\rm I}$ - was simple, albeit wrong. The problem of gauge coupling unification of Standard Model forces is the fact that $\alpha_1$ hits $\alpha_2$ too early, much below the lower limit on the GUT scale from proton longevity - and the only way to slow down the rise of $\alpha_1$ is to embed the $U(1)$ gauge group into a non-Abelian subgroup. 

As discussed in the previous section, typical values ensuring gauge coupling unification require $M_{\rm I}\sim 10^{11}\,\rm GeV$. However, as \eqref{nueffect} shows clearly, it fails in this minimal version of the theory with small representations - it produces too big neutrino mass. We have an indirect GERDA limit on neutrino Majorana mass from neutrinoless double beta decay $m_\nu \lesssim 0.2\, {\rm eV}$~\cite{GERDA:2020xhi}, but is obscured by flavor mixings - unlike the KATRIN direct limit  $m_\nu \lesssim {\rm eV}$~\cite{KATRIN:2021uub} from the endpoint of beta decay.

The end result is surprising: the intermediate scale $M_{\rm I}$ must be huge, close to the GUT scale, but that kills gauge coupling unification - unless there are some new light states. There must be an oasis in the desert, against the long standing prejudice. 
Notice also from \eqref{nueffect} that the cutoff must be as low as possible, which further strengthens the case for \eqref{Nmass} as the main source of N mass. But with  $\Lambda \gtrsim 10\, M_{\rm GUT}$, we are told that the GUT scale should be as low as possible, making the case for the potentially observable proton decay. 

The question is, what is lower bound on proton lifetime? As mentioned in Sec.~\ref{su5}, normally one argues that the bound $\tau_p \gtrsim 10^{34} \,{\rm yr}$~\cite{Super-Kamiokande:2020wjk} implies $M_{\rm GUT} \gtrsim 4 \cdot 10^{15}\, {\rm GeV}$. This is true under the condition that there are no judicial cancellation in proton decay amplitudes due to flavour conspiracies. In what follows, we will assume this natural scenario.

\begin{figure}[t]
    \centering
    \includegraphics[width=0.7\textwidth]{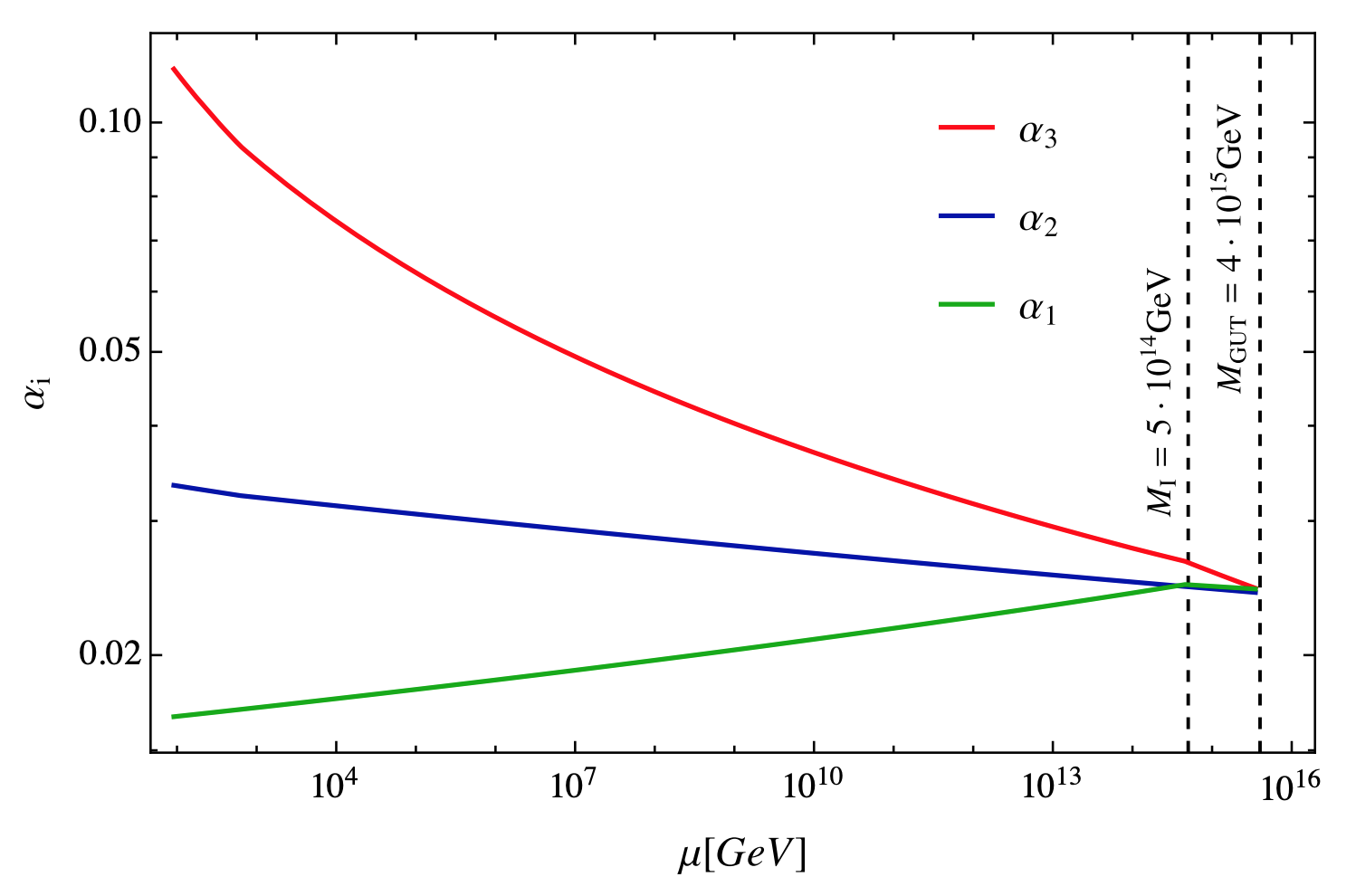}
    \caption{Explicit example of unification. Scalar weak triplet, scalar coloured octet and the scalar leptoquark are around $10\rm TeV$. Other particle thresholds are fixed to obtain unification. They can all be much heavier.}
    \label{fig:so10unif}
\end{figure}

\textit{Unification Constraints }
Taking into account neutrino-mass constraint of the previous section, resulting in $M_{\rm I}\sim M_{\rm GUT}$, the necessity of light particle states can be intuitively understood from the following 1-loop unification condition
\begin{equation}
\frac{M_{\rm GUT}}{M_Z} \simeq\exp{\left\{ \frac{\pi}{20}\left(5\alpha_{1}^{-1}-3\alpha_{2}^{-1}-2\alpha_{3}^{-1}\right)_{M_{Z}}\right\}}
\left(\frac{M_{Z}^4}{m_{3}\,m_{8}\,m_{\rm LQ}^2}\right)^{\frac{1}{40}},
\label{mainconstraint}
\end{equation}
where $m_3$ and $m_8$ are the masses of the scalar weak triplet and the scalar coloured octet. LQ denotes the leptoquark with quantum numbers $(3_C,2_W,1/6_Y)$ which descends from both $16_{\rm H}$ and $45_{\rm H}$. The requirement $M_{\rm GUT}\gtrsim 4\cdot 10^{15}\,\rm GeV$ implies, from \eqref{mainconstraint}, those new particle states to lie close to today's energies. This is analoguos to the $SU(5)$ situation c.f., Eqs.~(\ref{eq:mgutepsilon},\ref{eq:mgut24f}). 

To support our claim, a 2-loop renormalization group analysis was performed varying arbitrarily all particle thresholds in the spectrum. Also, thresholds effects due to gauge operators of the form 
\begin{equation}\label{kinetic}
{\rm Tr} \, F_{\mu \nu}   \left(\frac {\langle 45_{\rm H} \rangle}{\Lambda}\right)^n F^{\mu \nu}
\end{equation}
were taken into account. It should be noted that the $n=1$ operator vanishes due to the symmetry property of the operator. Therefore the leading correction comes from the dimension 6 operator. 

The end result is striking: The weak triplet, the coloured octet and the leptoquark all need to lie below $10\,\rm TeV$ in order to ensure gauge coupling unification and an intermediate mass scale compatible with observations. For more details on the phenomenology of these states see~\cite{Preda:2022izo,PSZ}. An explicit realization of the running is shown in Fig.~\ref{fig:so10unif}.

\subsection*{Renormalisable SO(10): $W_{\rm R}$ at colliders?}\label{WR}

The Left-Right symmetric subgroup $SU(3)_{\rm C}\times SU(2)_{\rm L}\times SU(2)_{\rm R}\times U(1)_{\rm B-L}$~\cite{Pati:1974yy,Mohapatra:1974gc,Senjanovic:1975rk,Senjanovic:1978ev}
of $SO(10)$ is of particular phenomenological interest due to the role it plays for neutrino mass and the fact that the scale $M_{W_{\rm R}}$ of its breaking may be accessible at future colliders, possibly even at the LHC. 
In this case, one could observe directly lepton number violation in the form of same charge di-lepton events accompanied with jets. This, so called Keung-Senjanovi\'c process~\cite{Keung:1983uu}, implies the same rate for opposite and same sign charged leptons and thus offer a unique direct probe the Majorana nature of RH neutrinos. Moreover, the theory offers a deep connection between low and high energy lepton number violation~\cite{Tello:2010am}, and allows for a self-contained, predictive origin of neutrino mass~\cite{Nemevsek:2012iq,Senjanovic:2016vxw,Senjanovic:2018xtu,Senjanovic:2019moe}. One can also predict the right-handed quark mixing matrix in terms of the usual left-handed CKM matrix~\cite{Senjanovic:2014pva,Senjanovic:2015yea} (for reviews of the LR symmetric model, the reader can consult~\cite{Senjanovic:2011zz,Tello:2012qda}, and a pedagogical {\it expose} can be found in~\cite{Melfo:2021wry}).
Thus, it is of great importance to know whether $M_{W_{\rm R}}$ could lie close to today's energies.

The question may sound surprising since we have seen above that the scale of $SU(2)_{\rm R}$ had to be basically the GUT scale. What happens, though, if one abandons the approach of small representations, and instead adheres to a renormalisable version of the theory? This requires, as is well know, a $126_{\rm H}$ dimensional Higgs representation needed to generate right-handed neutrino mass $m_{\rm N} \propto M_{\rm I}$. This can also correct the charged fermion mass relations~\cite{Babu:1992ia}. The minimality further requires to use the $45_{\rm H}$ adjoint representation to break the original symmetry at high energies~\cite{Bertolini:2009es}, together with the complex $10_{\rm H}$ field needed for the Standard Model symmetry breaking. So, the only change from the above minimal model is just the exchange of $16_{\rm H}$ with $126_{\rm H}$. This model has been re-discussed in recent years~\cite{Bertolini:2012im,Bertolini:2012az}, with a focus on the mass scales of the theory.

In order to perform the unification study, we must take into account that the Higgs scalars pertaining to the 
Left-Right symmetric model consist of the complex bi-doublet residing in $10_{\rm H} $ and $SU(2)_{\rm L,R}$ triplets~\cite{Minkowski:1977sc,Mohapatra:1979ia} (for details, see e.g.~\cite{Maiezza:2016ybz}) with $B-L=2$ from the $126_{\rm H}$. These fields must lie not above $M_{W_{\rm R}}$ due to the protective $SU(2)_{\rm L}\times SU(2)_{\rm R}\times U(1)_{\rm B-L}$ symmetry (the left-handed triplet is protected by the generalized charge conjugation $C$ residing in $SO(10)$, since $\langle 45_{\rm H}\rangle$ is invariant under it). 
A dedicated study~\cite{PSZ}, addressing the full freedom of the particle mass spectra, shows that in this case $M_{W_{\rm R}} \propto M_{\rm I} \propto \langle 126_{\rm H}\rangle$ could be arbitrary due to the proliferation of scalar states in $126_{\rm H}$ - and could even lie at today's energies.

This is unexpected since it has been a common lore that in the minimal renormalisable model $W_{\rm R}$-mass should lie at very high energies~\cite{delAguila:1980qag,Rizzo:1981su}, on the order of $10^{11}\,\rm GeV$ as mentioned above. The point is that this prejudice was based on the so-called extended survival principle according to which the scalar particle masses go to the largest possible value allowed by the symmetries in question~\cite{Mohapatra:1982aq}. This principle, however, fails completely as we saw in the above small Higgs representations scenarios. Once abandoned, all hell breaks loose when one deals with large representations and in that sense, the light $W_{\rm R}$ possibility appears fairly natural. Simply, some scalar states violate the survival principle - e.g., the Left-Right scale around $10^{4}\,\rm GeV$ can be obtained with~\cite{PSZ}: $(1_{\rm C},2_{\rm W}, 1/2_{\rm Y})_{126}$ at $10^{6}\,\rm GeV$, $(8_{\rm C},1_{\rm W}, 0_{\rm Y})_{45}$ at $10^8\,\rm GeV$, and $(3_{\rm C},1_{\rm W}, -2/3;1/3;4/3_{\rm Y})_{126}$ at $10^{12}\,\rm GeV$, where the subscript indicates the origin of the submultiplet. It should be stressed  that in the analysis both the scalar sector of the minimal Left-Right model as well as the tree-level mass sum rules of the spectrum~\cite{Bertolini:2012im}\footnote{These are even less constraining when taking into account radiative corrections~\cite{Bertolini:2009es,Bertolini:2012im}.} have been accounted for (all other masses are above $10^{14}\,\rm GeV$) in the running.

The point, however, is that all this happens due to the complete lack of predictivity and so the concept of grand unification becomes much less appealing: As shown above, the spectrum can easily be arranged in a way such that no new particles - apart from the ones realizing the low-energy Left-Right model - are close to present-day energies. This, once again, is due to a large proliferation of particle states. Moreover, the unification scale can be as high as $10^{17}\,\rm GeV$, thus making for a grim scenario regarding proton decay detections. Yet another pitfall, bringing the model to its dawn.
Still, a discovery of a light $W_{\rm R}$ would not discredit (but also not hint at) the minimal renormalizable $SO(10)$ theory, as often claimed in the literature. 

One can similarly ask whether the quark-lepton unification scale can be close to its lower phenomenological bound, which turns out be roughly $10^{5}$ GeV~\cite{Dolan:2020doe}. Since this is above present day or near future direct experimental reach, and since we are focused here on collider consequences of grand unification, we omit this discussion. Suffice it to say that the answer even in this case is positive~\footnote{It was claimed long ago~\cite{Senjanovic:1982ex} - in the context of extended survival principle - that LQ scale could be low, but that required much larger weak mixing angle, today known to be wrong}. 

There is even more to it. It turns out that the theory allows furthermore for observable neutrinoless double beta 
decay~\cite{Dvali:2023snt} and  $n-\bar n$ oscillations~\cite{Mohapatra:1980qe} dominated by new scalar states from the $\Delta_{\rm L,R}$
fields. 
An interested reader can find the details of this section in~\cite{PSZ}.

\section{Conclusion and Outlook}
\label{summary}

The idea of grand unification was received from the outset with great enthusiasm, and rightly so, for it predicted on general grounds the decay of the proton and the existence of magnetic monopoles, tied to charge quantization. There appeared to be a drawback, though, of this exciting avenue for beyond the Standard Model physics: it seemed to lead to a a desert in energies from the weak to the unification scale, with a bleak outcome of no directly observable physics for centuries. As the time went on, the physical couplings were measured better, and  it became clear that there must be an oasis (or more) in the desert. One possibility was low-energy supersymmetry~\cite{Dimopoulos:1981dw,Einhorn:1981sx,Marciano:1981un} with its protective mechanism for the hierarchy issue. However, this was not a must, only a possibility - an oasis could have been at a huge intermediate (not a good name really) as it seemed to happen generically in the $SO(10)$ theory~\cite{delAguila:1980qag,Rizzo:1981su} which, moreover, would be tied to the smallness of neutrino mass.

Some years ago it was noticed that, under the assumption of no cancellations in proton decay amplitudes, in the minimal realistic versions of the $SU(5)$ theory, an oasis had to be actually at rather low energies, potentially accessible even at the LHC~\cite{Bajc:2006ia}. And even more interesting, under equivalent assumption, the same turns out to be true even in the minimal $SO(10)$ theory with small representations~\cite{Preda:2022izo}, a great surprise. This is a game changer - all of a sudden grand unification could be tested directly, at least in some aspects, at hadron colliders. We have reviewed these models here and concentrated on one of the predicted light states: the real weak scalar triplet, which naturally gets tadpole-induced a small vacuum expectation value. Such a vacuum expectation value on order of few GeV, which fits perfectly with a particle at TeV energies, would then modify the Standard Model value of $W$-boson mass - and could easily account for the CDF anomaly~\cite{CDF:2022hxs}, if it were to be confirmed. 
We should stress that is of interest independently whether CDF is right or wrong - in the long run, if these models were correct one could expect a deviation from the Standard Model $W$-mass value.
Moreover, the low-energy effective theory of the triplet - due to grand unification constraints - turns out to be surprisingly predictive~\cite{Senjanovic:2022zwy}. Grand unified models with small representations thus offer new physics at low energies, naturally affecting the Standard Model predictions.

This comes at the price of higher-dimensional operators, though, which implies new physics beyond GUT. It is not surprising that many practitioners turned instead to large Higgs representations, which can work nicely at the renormalisable level. In the context of the $SU(5)$ theory, one needs to add a $45_{\rm H}$ representation that corrects wrong relations between down quark and charged lepton masses. We shied away from discussing this approach, since it brings a plethora of new states that prevent the predictiviy, without adding any new exciting physics.  

In the $SO(10)$ theory, however, this situation is more interesting. It amounts to trading $16_{\rm H}$ for $126_{\rm H}$, which can account for both the right-handed neutrino masses and correct the charged fermion mass relations. What about the mass scales in this model? There is no clear answer due to the proliferation of scalar states which obscure the unification constraints. 
The central question to ask then is how low can the intermediate scale of the Left-Right symmetry breaking be. When one ignores the threshold affects and simply assumes that the scalar masses take the largest possible values in accord with the symmetries in question, it turns out that $M_{\rm R}$ must be huge, far out in  the desert. This has led to the claim that $M_{\rm R}$ cannot lie at accessible energies. Allowing, however, for the physical freedom of scalar masses, implies that this is wrong - one can obtain easily even $M_{\rm R} \simeq \rm TeV$~\cite{PSZ}. Therefore, when sticking to renormalisable models predictions are lost. However, one gains the exciting possibility of new low-energy gauge dynamics.

{\bf Acknowledgements}  We wish to acknowledge the collaboration with Anca Preda on the $SO(10)$ theory. G.S. wishes to thank the organisers of the Corfu meeting for an invitation to present the ideas discussed here.

\printbibliography

@article{Georgi:1974yf,
    author = "Georgi, H. and Quinn, Helen R. and Weinberg, Steven",
    title = "{Hierarchy of Interactions in Unified Gauge Theories}",
    reportNumber = "Print-74-1122 Rev. (HARVARD), PRINT-74-1122 (HARVARD)",
    doi = "10.1103/PhysRevLett.33.451",
    journal = "Phys. Rev. Lett.",
    volume = "33",
    pages = "451--454",
    year = "1974"
}

@article{PSZ,
    author = "Preda, A. and Senjanovi\'c, G. and Zantedeschi, M.",
    eprint = "{in preparation}",
    journal = "{in preparation}"
}

@article{SZ,
    author = "Senjanovi\'c, G. and Zantedeschi, M.",
    eprint = "{in preparation}",
     journal = "{in preparation}"
}

@article{Bajc:2006ia,
    author = "Bajc, Borut and Senjanovi\'c, Goran",
    title = "{Seesaw at LHC}",
    eprint = "hep-ph/0612029",
    archivePrefix = "arXiv",
    doi = "10.1088/1126-6708/2007/08/014",
    journal = "JHEP",
    volume = "08",
    pages = "014",
    year = "2007"
}

@article{Georgi:1974sy,
    author = "Georgi, H. and Glashow, S. L.",
    title = "{Unity of All Elementary Particle Forces}",
    doi = "10.1103/PhysRevLett.32.438",
    journal = "Phys. Rev. Lett.",
    volume = "32",
    pages = "438--441",
    year = "1974"
}

@article{Fritzsch:1974nn,
    author = "Fritzsch, Harald and Minkowski, Peter",
    title = "{Unified Interactions of Leptons and Hadrons}",
    reportNumber = "CALT-68-467",
    doi = "10.1016/0003-4916(75)90211-0",
    journal = "Annals Phys.",
    volume = "93",
    pages = "193--266",
    year = "1975"
}

@article{Preda:2022izo,
    author = "Preda, Anca and Senjanovi\'c, Goran and Zantedeschi, Michael",
    title = "{SO(10): A case for hadron colliders}",
    eprint = "2201.02785",
    archivePrefix = "arXiv",
    primaryClass = "hep-ph",
    doi = "10.1016/j.physletb.2023.137746",
    journal = "Phys. Lett. B",
    volume = "838",
    pages = "137746",
    year = "2023"
}

@article{Minkowski:1977sc,
    author = "Minkowski, Peter",
    title = "{$\mu \to e\gamma$ at a Rate of One Out of $10^{9}$ Muon Decays?}",
    reportNumber = "Print-77-0182 (BERN)",
    doi = "10.1016/0370-2693(77)90435-X",
    journal = "Phys. Lett. B",
    volume = "67",
    pages = "421--428",
    year = "1977"
}

@article{Mohapatra:1979ia,
    author = "Mohapatra, Rabindra N. and Senjanovi\'c, Goran",
    title = "{Neutrino Mass and Spontaneous Parity Nonconservation}",
    reportNumber = "MDDP-TR-80-060, MDDP-PP-80-105, CCNY-HEP-79-10",
    doi = "10.1103/PhysRevLett.44.912",
    journal = "Phys. Rev. Lett.",
    volume = "44",
    pages = "912",
    year = "1980"
}

@article{GellMann:1980vs,
    author = "Gell-Mann, Murray and Ramond, Pierre and Slansky, Richard",
    title = "{Complex Spinors and Unified Theories}",
    eprint = "1306.4669",
    archivePrefix = "arXiv",
    primaryClass = "hep-th",
    reportNumber = "PRINT-80-0576",
    journal = "Conf. Proc. C",
    volume = "790927",
    pages = "315--321",
    year = "1979"
}

@article{Glashow:1979nm,
    author = "Glashow, S. L.",
    title = "{The Future of Elementary Particle Physics}",
    reportNumber = "HUTP-79-A059",
    doi = "10.1007/978-1-4684-7197-7_15",
    journal = "NATO Sci. Ser. B",
    volume = "61",
    pages = "687",
    year = "1980"
}

@article{Bajc:2015ita,
    author = "Bajc, Borut and Lavignac, St\'ephane and Mede, Timon",
    title = "{Resurrecting the minimal renormalizable supersymmetric SU(5) model}",
    eprint = "1509.06680",
    archivePrefix = "arXiv",
    primaryClass = "hep-ph",
    doi = "10.1007/JHEP01(2016)044",
    journal = "JHEP",
    volume = "01",
    pages = "044",
    year = "2016"
}

@article{Dimopoulos:1981dw,
    author = "Dimopoulos, Savas and Raby, Stuart and Wilczek, Frank",
    title = "{Proton Decay in Supersymmetric Models}",
    reportNumber = "NSF-ITP-82-08, UM HE 81-64",
    doi = "10.1016/0370-2693(82)90313-6",
    journal = "Phys. Lett. B",
    volume = "112",
    pages = "133",
    year = "1982"
}

@article{Yanagida:1979as,
    author = "Yanagida, Tsutomu",
    editor = "Sawada, Osamu and Sugamoto, Akio",
    title = "{Horizontal gauge symmetry and masses of neutrinos}",
    reportNumber = "KEK-79-18-95",
    journal = "Conf. Proc. C",
    volume = "7902131",
    pages = "95--99",
    year = "1979"
}

@article{Senjanovic:2022zwy,
    author = "Senjanovi\'c, Goran and Zantedeschi, Michael",
    title = "{SU(5) grand unification and W-boson mass}",
    eprint = "2205.05022",
    archivePrefix = "arXiv",
    primaryClass = "hep-ph",
    doi = "10.1016/j.physletb.2022.137653",
    journal = "Phys. Lett. B",
    volume = "837",
    pages = "137653",
    year = "2023"
}

@article{CDF:2022hxs,
    author = "Aaltonen, T. and others",
    collaboration = "CDF",
    title = "{High-precision measurement of the $W$          boson mass with the CDF II detector}",
    reportNumber = "FERMILAB-PUB-22-254-PPD",
    doi = "10.1126/science.abk1781",
    journal = "Science",
    volume = "376",
    number = "6589",
    pages = "170--176",
    year = "2022"
}

@article{Einhorn:1981sx,
    author = "Einhorn, M. B. and Jones, D. R. T.",
    title = "{The Weak Mixing Angle and Unification Mass in Supersymmetric SU(5)}",
    reportNumber = "UM HE 81-55",
    doi = "10.1016/0550-3213(82)90502-8",
    journal = "Nucl. Phys. B",
    volume = "196",
    pages = "475--488",
    year = "1982"
}

@article{Marciano:1981un,
    author = "Marciano, William J. and Senjanovi\'c, Goran",
    title = "{Predictions of Supersymmetric Grand Unified Theories}",
    reportNumber = "Print-81-0844 (BROOKHAVEN), BNL-30398",
    doi = "10.1103/PhysRevD.25.3092",
    journal = "Phys. Rev. D",
    volume = "25",
    pages = "3092",
    year = "1982"
}

@article{Senjanovic:2011zz,
    author = "Senjanovi\'c, G.",
    title = "{Neutrino mass: From LHC to grand unification}",
    doi = "10.1393/ncr/i2011-10061-8",
    journal = "Riv. Nuovo Cim.",
    volume = "34",
    number = "1",
    pages = "1--68",
    year = "2011"
}

@article{Super-Kamiokande:2020wjk,
    author = "Takenaka, A. and others",
    collaboration = "Super-Kamiokande",
    title = "{Search for proton decay via $p\to e^+\pi^0$ and $p\to \mu^+\pi^0$ with an enlarged fiducial volume in Super-Kamiokande I-IV}",
    eprint = "2010.16098",
    archivePrefix = "arXiv",
    primaryClass = "hep-ex",
    doi = "10.1103/PhysRevD.102.112011",
    journal = "Phys. Rev. D",
    volume = "102",
    number = "11",
    pages = "112011",
    year = "2020"
}

@article{Nandi:1982ew,
    author = "Nandi, S. and Stern, A. and Sudarshan, E. C. G.",
    title = "{CAN PROTON DECAY BE ROTATED AWAY?}",
    reportNumber = "DOE-ER-03992-472",
    doi = "10.1016/0370-2693(82)90416-6",
    journal = "Phys. Lett. B",
    volume = "113",
    pages = "165--169",
    year = "1982"
}

@article{Dorsner:2004xa,
    author = "Dorsner, Ilja and Fileviez Perez, Pavel",
    title = "{How long could we live?}",
    eprint = "hep-ph/0410198",
    archivePrefix = "arXiv",
    doi = "10.1016/j.physletb.2005.08.039",
    journal = "Phys. Lett. B",
    volume = "625",
    pages = "88--95",
    year = "2005"
}

@article{Shafi:1979qb,
    author = "Shafi, Q. and Wetterich, C.",
    title = "{Gauge Hierarchies and the Unification Mass}",
    reportNumber = "CERN-TH-2667",
    doi = "10.1016/0370-2693(79)90775-5",
    journal = "Phys. Lett. B",
    volume = "85",
    pages = "52--56",
    year = "1979"
}

@article{Weinberg:1979sa,
    author = "Weinberg, Steven",
    title = "{Baryon and Lepton Nonconserving Processes}",
    reportNumber = "HUTP-79-A050",
    doi = "10.1103/PhysRevLett.43.1566",
    journal = "Phys. Rev. Lett.",
    volume = "43",
    pages = "1566--1570",
    year = "1979"
}

@article{Magg:1980ut,
    author = "Magg, M. and Wetterich, C.",
    title = "{Neutrino Mass Problem and Gauge Hierarchy}",
    reportNumber = "CERN-TH-2829",
    doi = "10.1016/0370-2693(80)90825-4",
    journal = "Phys. Lett. B",
    volume = "94",
    pages = "61--64",
    year = "1980"
}

@article{Mohapatra:1980yp,
    author = "Mohapatra, Rabindra N. and Senjanovi\'c, Goran",
    title = "{Neutrino Masses and Mixings in Gauge Models with Spontaneous Parity Violation}",
    reportNumber = "FERMILAB-PUB-80-061-THY, FERMILAB-PUB-80-061-T",
    doi = "10.1103/PhysRevD.23.165",
    journal = "Phys. Rev. D",
    volume = "23",
    pages = "165",
    year = "1981"
}

@article{Lazarides:1980nt,
    author = "Lazarides, George and Shafi, Q. and Wetterich, C.",
    title = "{Proton Lifetime and Fermion Masses in an SO(10) Model}",
    reportNumber = "FREIBURG-THEP-80-2",
    doi = "10.1016/0550-3213(81)90354-0",
    journal = "Nucl. Phys. B",
    volume = "181",
    pages = "287--300",
    year = "1981"
}

@article{Dorsner:2005fq,
    author = "Dorsner, Ilja and Fileviez Perez, Pavel",
    title = "{Unification without supersymmetry: Neutrino mass, proton decay and light leptoquarks}",
    eprint = "hep-ph/0504276",
    archivePrefix = "arXiv",
    doi = "10.1016/j.nuclphysb.2005.06.016",
    journal = "Nucl. Phys. B",
    volume = "723",
    pages = "53--76",
    year = "2005"
}

@article{Garayoa:2007fw,
    author = "Garayoa, Julia and Schwetz, Thomas",
    title = "{Neutrino mass hierarchy and Majorana CP phases within the Higgs triplet model at the LHC}",
    eprint = "0712.1453",
    archivePrefix = "arXiv",
    primaryClass = "hep-ph",
    reportNumber = "CERN-PH-TH-2007-255, IFIC-07-75, FTUV-07-1210",
    doi = "10.1088/1126-6708/2008/03/009",
    journal = "JHEP",
    volume = "03",
    pages = "009",
    year = "2008"
}

@article{Foot:1988aq,
    author = "Foot, Robert and Lew, H. and He, X. G. and Joshi, Girish C.",
    title = "{Seesaw Neutrino Masses Induced by a Triplet of Leptons}",
    reportNumber = "UM-P-88/89, OZ-P-88/7",
    doi = "10.1007/BF01415558",
    journal = "Z. Phys. C",
    volume = "44",
    pages = "441",
    year = "1989"
}

@article{Bajc:2007zf,
    author = "Bajc, Borut and Nemevsek, Miha and Senjanovi\'c, Goran",
    title = "{Probing seesaw at LHC}",
    eprint = "hep-ph/0703080",
    archivePrefix = "arXiv",
    doi = "10.1103/PhysRevD.76.055011",
    journal = "Phys. Rev. D",
    volume = "76",
    pages = "055011",
    year = "2007"
}

@article{Arhrib:2009mz,
    author = "Arhrib, Abdesslam and Bajc, Borut and Ghosh, Dilip Kumar and Han, Tao and Huang, Gui-Yu and Puljak, Ivica and Senjanovi\'c, Goran",
    title = "{Collider Signatures for Heavy Lepton Triplet in Type I+III Seesaw}",
    eprint = "0904.2390",
    archivePrefix = "arXiv",
    primaryClass = "hep-ph",
    reportNumber = "MADPH-08-1526",
    doi = "10.1103/PhysRevD.82.053004",
    journal = "Phys. Rev. D",
    volume = "82",
    pages = "053004",
    year = "2010"
}

@phdthesis{Novak:2020jju,
    author = "Novak, Tadej",
    title = "{Search for type-III seesaw heavy leptons with the ATLAS detector at the LHC}",
    school = "Ljubljana U.",
    year = "2020"
}

@article{Chiang:2020rcv,
    author = "Chiang, Cheng-Wei and Cottin, Giovanna and Du, Yong and Fuyuto, Kaori and Ramsey-Musolf, Michael J.",
    title = "{Collider Probes of Real Triplet Scalar Dark Matter}",
    eprint = "2003.07867",
    archivePrefix = "arXiv",
    primaryClass = "hep-ph",
    doi = "10.1007/JHEP01(2021)198",
    journal = "JHEP",
    volume = "01",
    pages = "198",
    year = "2021"
}

@article{Buras:1977yy,
    author = "Buras, A. J. and Ellis, John R. and Gaillard, M. K. and Nanopoulos, Dimitri V.",
    title = "{Aspects of the Grand Unification of Strong, Weak and Electromagnetic Interactions}",
    reportNumber = "CERN-TH-2403",
    doi = "10.1016/0550-3213(78)90214-6",
    journal = "Nucl. Phys. B",
    volume = "135",
    pages = "66--92",
    year = "1978"
}

@article{ParticleDataGroup:2020ssz,
    author = "Zyla, P. A. and others",
    collaboration = "Particle Data Group",
    title = "{Review of Particle Physics}",
    doi = "10.1093/ptep/ptaa104",
    journal = "PTEP",
    volume = "2020",
    number = "8",
    pages = "083C01",
    year = "2020"
}

@article{Cheng:2022jyi,
    author = "Cheng, Yu and He, Xiao-Gang and Huang, Zhong-Lv and Li, Ming-Wei",
    title = "{Type-II seesaw triplet scalar effects on neutrino trident scattering}",
    eprint = "2204.05031",
    archivePrefix = "arXiv",
    primaryClass = "hep-ph",
    doi = "10.1016/j.physletb.2022.137218",
    journal = "Phys. Lett. B",
    volume = "831",
    pages = "137218",
    year = "2022"
}

@article{Heeck:2022fvl,
    author = "Heeck, Julian",
    title = "{W-boson mass in the triplet seesaw model}",
    eprint = "2204.10274",
    archivePrefix = "arXiv",
    primaryClass = "hep-ph",
    doi = "10.1103/PhysRevD.106.015004",
    journal = "Phys. Rev. D",
    volume = "106",
    number = "1",
    pages = "015004",
    year = "2022"
}

@article{tHooft:1979rat,
    author = "'t Hooft, Gerard",
    editor = "'t Hooft, Gerard and Itzykson, C. and Jaffe, A. and Lehmann, H. and Mitter, P. K. and Singer, I. M. and Stora, R.",
    title = "{Naturalness, chiral symmetry, and spontaneous chiral symmetry breaking}",
    reportNumber = "PRINT-80-0083 (UTRECHT)",
    doi = "10.1007/978-1-4684-7571-5_9",
    journal = "NATO Sci. Ser. B",
    volume = "59",
    pages = "135--157",
    year = "1980"
}

@article{Dimopoulos:1981zb,
    author = "Dimopoulos, Savas and Georgi, Howard",
    title = "{Softly Broken Supersymmetry and SU(5)}",
    reportNumber = "HUTP-81/A022",
    doi = "10.1016/0550-3213(81)90522-8",
    journal = "Nucl. Phys. B",
    volume = "193",
    pages = "150--162",
    year = "1981"
}

@article{Arkani-Hamed:2004ymt,
    author = "Arkani-Hamed, Nima and Dimopoulos, Savas",
    title = "{Supersymmetric unification without low energy supersymmetry and signatures for fine-tuning at the LHC}",
    eprint = "hep-th/0405159",
    archivePrefix = "arXiv",
    doi = "10.1088/1126-6708/2005/06/073",
    journal = "JHEP",
    volume = "06",
    pages = "073",
    year = "2005"
}

@article{Pati:1974yy,
    author = "Pati, Jogesh C. and Salam, Abdus",
    title = "{Lepton Number as the Fourth Color}",
    reportNumber = "IC-74-7",
    doi = "10.1103/PhysRevD.10.275",
    journal = "Phys. Rev. D",
    volume = "10",
    pages = "275--289",
    year = "1974",
    note = "[Erratum: Phys.Rev.D 11, 703--703 (1975)]"
}

@article{delAguila:1980qag,
    author = "del Aguila, F. and Ibanez, Luis E.",
    title = "{Higgs Bosons in SO(10) and Partial Unification}",
    reportNumber = "OXFORD-TP 41/80",
    doi = "10.1016/0550-3213(81)90266-2",
    journal = "Nucl. Phys. B",
    volume = "177",
    pages = "60--86",
    year = "1981"
}

@article{Rizzo:1981su,
    author = "Rizzo, Thomas G. and Senjanovi\'c, Goran",
    title = "{Can There Be Low Intermediate Mass Scales in Grand Unified Theories?}",
    reportNumber = "BNL-28992",
    doi = "10.1103/PhysRevLett.46.1315",
    journal = "Phys. Rev. Lett.",
    volume = "46",
    pages = "1315",
    year = "1981"
}

@article{Witten:1979nr,
    author = "Witten, Edward",
    title = "{Neutrino Masses in the Minimal O(10) Theory}",
    reportNumber = "HUTP-79/A076",
    doi = "10.1016/0370-2693(80)90666-8",
    journal = "Phys. Lett. B",
    volume = "91",
    pages = "81--84",
    year = "1980"
}

@article{Dvali:2007hz,
    author = "Dvali, Gia",
    title = "{Black Holes and Large N Species Solution to the Hierarchy Problem}",
    eprint = "0706.2050",
    archivePrefix = "arXiv",
    primaryClass = "hep-th",
    doi = "10.1002/prop.201000009",
    journal = "Fortsch. Phys.",
    volume = "58",
    pages = "528--536",
    year = "2010"
}

@article{Dvali:2007wp,
    author = "Dvali, Gia and Redi, Michele",
    title = "{Black Hole Bound on the Number of Species and Quantum Gravity at LHC}",
    eprint = "0710.4344",
    archivePrefix = "arXiv",
    primaryClass = "hep-th",
    doi = "10.1103/PhysRevD.77.045027",
    journal = "Phys. Rev. D",
    volume = "77",
    pages = "045027",
    year = "2008"
}

@article{GERDA:2020xhi,
    author = "Agostini, M. and others",
    collaboration = "GERDA",
    title = "{Final Results of GERDA on the Search for Neutrinoless Double-$\beta$ Decay}",
    eprint = "2009.06079",
    archivePrefix = "arXiv",
    primaryClass = "nucl-ex",
    doi = "10.1103/PhysRevLett.125.252502",
    journal = "Phys. Rev. Lett.",
    volume = "125",
    number = "25",
    pages = "252502",
    year = "2020"
}

@article{KATRIN:2021uub,
    author = "Aker, M. and others",
    collaboration = "KATRIN",
    title = "{Direct neutrino-mass measurement with sub-electronvolt sensitivity}",
    eprint = "2105.08533",
    archivePrefix = "arXiv",
    primaryClass = "hep-ex",
    doi = "10.1038/s41567-021-01463-1",
    journal = "Nature Phys.",
    volume = "18",
    number = "2",
    pages = "160--166",
    year = "2022"
}

@article{Mohapatra:1974gc,
    author = "Mohapatra, R. N. and Pati, Jogesh C.",
    title = "{A Natural Left-Right Symmetry}",
    reportNumber = "CCNY-HEP-74-2",
    doi = "10.1103/PhysRevD.11.2558",
    journal = "Phys. Rev. D",
    volume = "11",
    pages = "2558",
    year = "1975"
}

@article{Senjanovic:1975rk,
    author = "Senjanovi\'c, G. and Mohapatra, Rabindra N.",
    title = "{Exact Left-Right Symmetry and Spontaneous Violation of Parity}",
    reportNumber = "CCNY-HEP-75-5",
    doi = "10.1103/PhysRevD.12.1502",
    journal = "Phys. Rev. D",
    volume = "12",
    pages = "1502",
    year = "1975"
}

@article{Senjanovic:1978ev,
    author = "Senjanovi\'c, Goran",
    title = "{Spontaneous Breakdown of Parity in a Class of Gauge Theories}",
    reportNumber = "CCNY-HEP-78/20",
    doi = "10.1016/0550-3213(79)90604-7",
    journal = "Nucl. Phys. B",
    volume = "153",
    pages = "334--364",
    year = "1979"
}

@article{Keung:1983uu,
    author = "Keung, Wai-Yee and Senjanovi\'c, Goran",
    title = "{Majorana Neutrinos and the Production of the Right-handed Charged Gauge Boson}",
    reportNumber = "BNL-32872",
    doi = "10.1103/PhysRevLett.50.1427",
    journal = "Phys. Rev. Lett.",
    volume = "50",
    pages = "1427",
    year = "1983"
}

@article{Tello:2010am,
    author = "Tello, Vladimir and Nemevsek, Miha and Nesti, Fabrizio and Senjanovi\'c, Goran and Vissani, Francesco",
    title = "{Left-Right Symmetry: from LHC to Neutrinoless Double Beta Decay}",
    eprint = "1011.3522",
    archivePrefix = "arXiv",
    primaryClass = "hep-ph",
    doi = "10.1103/PhysRevLett.106.151801",
    journal = "Phys. Rev. Lett.",
    volume = "106",
    pages = "151801",
    year = "2011"
}

@article{Nemevsek:2012iq,
    author = "Nemevsek, Miha and Senjanovi\'c, Goran and Tello, Vladimir",
    title = "{Connecting Dirac and Majorana Neutrino Mass Matrices in the Minimal Left-Right Symmetric Model}",
    eprint = "1211.2837",
    archivePrefix = "arXiv",
    primaryClass = "hep-ph",
    doi = "10.1103/PhysRevLett.110.151802",
    journal = "Phys. Rev. Lett.",
    volume = "110",
    number = "15",
    pages = "151802",
    year = "2013"
}

@article{Senjanovic:2016vxw,
    author = "Senjanovi\'c, Goran and Tello, Vladimir",
    title = "{Probing Seesaw with Parity Restoration}",
    eprint = "1612.05503",
    archivePrefix = "arXiv",
    primaryClass = "hep-ph",
    doi = "10.1103/PhysRevLett.119.201803",
    journal = "Phys. Rev. Lett.",
    volume = "119",
    number = "20",
    pages = "201803",
    year = "2017"
}

@article{Senjanovic:2018xtu,
    author = "Senjanovi\'c, Goran and Tello, Vladimir",
    title = "{Disentangling the seesaw mechanism in the minimal left-right symmetric model}",
    eprint = "1812.03790",
    archivePrefix = "arXiv",
    primaryClass = "hep-ph",
    doi = "10.1103/PhysRevD.100.115031",
    journal = "Phys. Rev. D",
    volume = "100",
    number = "11",
    pages = "115031",
    year = "2019"
}

@article{Senjanovic:2019moe,
    author = "Senjanovi\'c, Goran and Tello, Vladimir",
    title = "{Parity and the origin of neutrino mass}",
    eprint = "1912.13060",
    archivePrefix = "arXiv",
    primaryClass = "hep-ph",
    doi = "10.1142/S0217751X20500530",
    journal = "Int. J. Mod. Phys. A",
    volume = "35",
    number = "09",
    pages = "2050053",
    year = "2020"
}

@article{Senjanovic:2014pva,
    author = "Senjanovi\'c, Goran and Tello, Vladimir",
    title = "{Right Handed Quark Mixing in Left-Right Symmetric Theory}",
    eprint = "1408.3835",
    archivePrefix = "arXiv",
    primaryClass = "hep-ph",
    doi = "10.1103/PhysRevLett.114.071801",
    journal = "Phys. Rev. Lett.",
    volume = "114",
    number = "7",
    pages = "071801",
    year = "2015"
}

@article{Senjanovic:2015yea,
    author = "Senjanovi\'c, Goran and Tello, Vladimir",
    title = "{Restoration of Parity and the Right-Handed Analog of the CKM Matrix}",
    eprint = "1502.05704",
    archivePrefix = "arXiv",
    primaryClass = "hep-ph",
    doi = "10.1103/PhysRevD.94.095023",
    journal = "Phys. Rev. D",
    volume = "94",
    number = "9",
    pages = "095023",
    year = "2016"
}

@phdthesis{Tello:2012qda,
    author = "Tello, Vladimir",
    title = "{Connections between the high and low energy violation of Lepton and Flavor numbers in the minimal left-right symmetric model}",
    school = "SISSA, Trieste",
    year = "2012"
}

@article{Melfo:2021wry,
    author = "Melfo, Alejandra and Senjanovi\'c, Goran",
    title = "{Neutrino: Chronicles of an aloof protagonist}",
    eprint = "2107.05472",
    archivePrefix = "arXiv",
    primaryClass = "physics.hist-ph",
    doi = "10.1142/S0217732322300087",
    journal = "Mod. Phys. Lett. A",
    volume = "37",
    number = "12",
    pages = "2230008",
    year = "2022"
}

@article{Babu:1992ia,
    author = "Babu, K. S. and Mohapatra, R. N.",
    title = "{Predictive neutrino spectrum in minimal SO(10) grand unification}",
    eprint = "hep-ph/9209215",
    archivePrefix = "arXiv",
    reportNumber = "BA-92-54, UMD-PP-93-021",
    doi = "10.1103/PhysRevLett.70.2845",
    journal = "Phys. Rev. Lett.",
    volume = "70",
    pages = "2845--2848",
    year = "1993"
}

@article{Bertolini:2009es,
    author = "Bertolini, Stefano and Di Luzio, Luca and Malinsky, Michal",
    title = "{On the vacuum of the minimal nonsupersymmetric SO(10) unification}",
    eprint = "0912.1796",
    archivePrefix = "arXiv",
    primaryClass = "hep-ph",
    doi = "10.1103/PhysRevD.81.035015",
    journal = "Phys. Rev. D",
    volume = "81",
    pages = "035015",
    year = "2010"
}

@article{Bertolini:2012im,
    author = "Bertolini, Stefano and Di Luzio, Luca and Malinsky, Michal",
    title = "{Seesaw Scale in the Minimal Renormalizable SO(10) Grand Unification}",
    eprint = "1202.0807",
    archivePrefix = "arXiv",
    primaryClass = "hep-ph",
    reportNumber = "IFIC-12-07, TTP12-004, SFB-CPP-12-06",
    doi = "10.1103/PhysRevD.85.095014",
    journal = "Phys. Rev. D",
    volume = "85",
    pages = "095014",
    year = "2012"
}

@article{Bertolini:2012az,
    author = "Bertolini, Stefano and Di Luzio, Luca and Malinsky, Michal",
    editor = "Fukuyama, Takeshi and Mohapatra, Rabindra Nath",
    title = "{Towards a New Minimal SO(10) Unification}",
    eprint = "1205.5637",
    archivePrefix = "arXiv",
    primaryClass = "hep-ph",
    doi = "10.1063/1.4742077",
    journal = "AIP Conf. Proc.",
    volume = "1467",
    pages = "37--44",
    year = "2012"
}

@article{Mohapatra:1980qe,
    author = "Mohapatra, Rabindra N. and Marshak, R. E.",
    title = "{Local B-L Symmetry of Electroweak Interactions, Majorana Neutrinos and Neutron Oscillations}",
    reportNumber = "VPI-HEP-80/1",
    doi = "10.1103/PhysRevLett.44.1316",
    journal = "Phys. Rev. Lett.",
    volume = "44",
    pages = "1316--1319",
    year = "1980",
    note = "[Erratum: Phys.Rev.Lett. 44, 1643 (1980)]"
}

@article{Dvali:2023snt,
    author = "Dvali, Gia and Maiezza, Alessio and Senjanovi\'c, Goran and Tello, Vladimir",
    title = "{Neutrinoless double beta decay as seen by the devil's advocate}",
    eprint = "2303.17261",
    archivePrefix = "arXiv",
    primaryClass = "hep-ph",
    month = "3",
    year = "2023"
}

@article{Maiezza:2016ybz,
    author = "Maiezza, Alessio and Senjanovi\'c, Goran and Vasquez, Juan Carlos",
    title = "{Higgs sector of the minimal left-right symmetric theory}",
    eprint = "1612.09146",
    archivePrefix = "arXiv",
    primaryClass = "hep-ph",
    doi = "10.1103/PhysRevD.95.095004",
    journal = "Phys. Rev. D",
    volume = "95",
    number = "9",
    pages = "095004",
    year = "2017"
}

@article{Mohapatra:1982aq,
    author = "Mohapatra, Rabindra N. and Senjanovi\'c, Goran",
    title = "{Higgs Boson Effects in Grand Unified Theories}",
    reportNumber = "CCNY-HEP-82/8a",
    doi = "10.1103/PhysRevD.27.1601",
    journal = "Phys. Rev. D",
    volume = "27",
    pages = "1601",
    year = "1983"
}

@article{Dolan:2020doe,
    author = "Dolan, Matthew J. and Dutka, Tomasz P. and Volkas, Raymond R.",
    title = "{Lowering the scale of Pati-Salam breaking through seesaw mixing}",
    eprint = "2012.05976",
    archivePrefix = "arXiv",
    primaryClass = "hep-ph",
    doi = "10.1007/JHEP05(2021)199",
    journal = "JHEP",
    volume = "05",
    pages = "199",
    year = "2021"
}

@article{Senjanovic:1982ex,
    author = "Senjanovi\'c, Goran and Sokorac, Aleksandar",
    title = "{Light Leptoquarks in SO(10)}",
    reportNumber = "Print-82-0652 (BNL), BNL-31824",
    doi = "10.1007/BF01574858",
    journal = "Z. Phys. C",
    volume = "20",
    pages = "255",
    year = "1983"
}
\end{document}